\documentclass[aps,prl,superscriptaddress,amsfonts,amssymb,amsmath,,notitlepage,11pt]{revtex4-1}
\usepackage{graphicx}  
\usepackage{bm}
\usepackage{color}
\usepackage{ulem} 
\usepackage[labelformat=empty]{caption}

\begin{document}

\title{Single-Site Non-Fermi Liquid Behaviors in a Diluted 4$f^2$ System Y$_{1-x}$Pr$_x$Ir$_2$Zn$_{20}$}



\author{Y. Yamane} \affiliation{Graduate School of Advanced Sciences of Matter, Hiroshima University, Higashi-Hiroshima 739-8530, Japan}
\author{T. Onimaru} \affiliation{Graduate School of Advanced Sciences of Matter, Hiroshima University, Higashi-Hiroshima 739-8530, Japan}
\author{K. Wakiya} \thanks{Present address: Faculty of Engineering, Yokohama National University, Yokohama 240-8501, Japan.} \affiliation{Graduate School of Advanced Sciences of Matter, Hiroshima University, Higashi-Hiroshima 739-8530, Japan}
\author{K. T. Matsumoto} \thanks{Present address: Graduate School of Science and Engineering, Ehime University, Matsuyama 790-8577, Japan} \affiliation{Graduate School of Advanced Sciences of Matter, Hiroshima University, Higashi-Hiroshima 739-8530, Japan}
\author{K. Umeo} \affiliation{Cryogenic and Instrumental Analysis Division, N-BARD, Hiroshima University, Higashi-Hiroshima 739-8526, Japan}
\author{T. Takabatake} \affiliation{Graduate School of Advanced Sciences of Matter, Hiroshima University, Higashi-Hiroshima 739-8530, Japan}


\date{\today}

\begin{abstract}
Electrical resistivity $\rho(T)$ and specific heat $C(T)$ measurements have been made on the diluted 4$f^2$ system Y(Pr)Ir$_2$Zn$_{20}$.
Both data of $\rho$ and magnetic specific heat $C_{\rm m}$ per Pr ion are well scaled as a function of $T/T_{\rm 0}$, where $T_{\rm 0}$ is a characteristic temperature of {non-Fermi liquid (NFL) behaviors}.
Furthermore, the temperature dependences of $\rho$ and $C_{\mathrm{m}}/T$ agree with the NFL behaviors predicted by the two-channel Kondo model for the strong coupling limit.
Therefore, we infer that the observed NFL behaviors result from the single-site quadrupole Kondo effect due to the hybridization of the 4$f^2$ states with multi-channel conduction electrons.
\end{abstract}


\maketitle

Low-temperature physical properties of correlated electronic systems usually follow the Landau Fermi-liquid (FL) theory.
When the FL state could become unstable, an anomalous metallic state involving a so-called non-Fermi-liquid (NFL) behavior would emerge \cite{Stewart_2001}.
The NFL phenomena in $f$-electron systems mostly accompany quantum criticality where the quantum fluctuations are divergently enhanced by strong competition between the on-site single-channel Kondo effect and the inter-site Ruderman-Kittel-Kasuya-Yosida interaction \cite{Gegenwart_2008}.
On the other hand, another type of NFL state was proposed by Nozi$\grave{\rm e}$res and Blandin, where an impurity spin is over-screened by spin and orbital degrees of freedom due to multiple channels of conduction bands, that is mutli-channel Kondo effect \cite{Nozieres_1980}.
Cox adapted it for a local electric quadrupole of $f^2$ configuration \cite{Cox_1987} to discribe NFL behaviors observed in a 5$f$-electron heavy fermion system UBe$_{13}$ \cite{Ott_1983}. 
The ground state of the 5$f^2$ configuration of U$^{4+}$ ion under the cubic crystalline electric field (CEF) was assumed to be the non-Kramers doublet with active quadrupoles.
The basic idea is that the local quadrupole is over-screened by two-channel conduction bands, where the degenerated magnetic moments of equivalent conduction bands play a role for the scattering channels.
This quadrupole Kondo effect leads to the NFL behaviors: specific heat divided by temperature $C/T \propto$ $-$ln$T$, quadrupole susceptibility $\chi_{\rm Q} \propto$ $-$ln$T$, and normalized resistivity $\rho/\rho_0 \propto 1+ A\sqrt{T}$, where $\rho_0$ is the residual value and $A$ is a coefficient \cite{Cox_1998}.
Another striking prediction is the existence of the residual entropy of $S = 0.5R$ln$2$ at zero temperature, where $R$ is the gas constant.
In recent years, the fractional entropy due to the over-screened quadrupoles has been focused in terms of fictitious Majorana fermions that could play key roles in two-dimensional superconductors and topological insulators \cite{Emery_1992, Alicea_2012}. 

Following the discovery of the NFL behaviors in UBe$_{13}$, similar behaviors were observed successively in diluted 5$f$ systems such as Th(U)Be$_{13}$, Th(U)Ru$_2$Si$_2$, and Y(U)Pd$_3$ \cite{Amitsuka_1994, Aliev_1994, Seaman_1991}.
However, it has been argued that the NFL behaviors could be explained by other mechanisms such as the competition between the CEF singlet and the Kondo-Yosida singlet \cite{Yotsuhashi_2002, Hattori_2005, Nishiyama_2010}, a magnetic instability due to the inter-site magnetic interaction \cite{Andraka_1991}, and hexadecapolar Kondo effect \cite{Toth_2011}.
Moreover, atomic disorder in the samples is inevitable and the CEF levels and valences of the U ions are not well determined.
These problems make it difficult to judge whether or not the NFL behaviors in these 5$f^2$ systems result from the quadrupole Kondo effect.

In Pr-based intermetallic systems with 4$f^2$ configuration, on the other hand, the prerequisite for the quadrupole Kondo effect can be fulfilled. 
In fact, a diluted Pr system of cubic La(Pr)Pb$_{3}$ exhibits the $-$ln$T$ behavior in $C/T$ \cite{Kawae_2006}.
However, the superconducting filament of Pb precipitating on the sample surface concealed the intrinsic behavior of $\rho(T)$ \cite{Kawae_2003}.
It is therefore necessary to find more suitable system for establishing the quadrupole Kondo effect.

Recently, a family of compounds Pr$T_2X_{20}$ ($T$ = transition metal, $X$ = Al, Zn, and Cd) have been intensively studied \cite{Onimaru16_JPSJ}.
They show a variety of phenomena such as long-range quadrupole order, unconventional superconductivity, and NFL behavior in which the quadrupoles of the non-Kramers doublet ground state of the Pr$^{3+}$ ion are involved. 

PrIr$_2$Zn$_{20}$, for instance, displays an antiferroquadrupolar (AFQ) order at $T_{\rm Q}$ $=$ 0.11 K \cite{Onimaru_2011,Ishii_2011,Iwasa_2017}, below which a superconducting transition occurs at $T_{\rm c}$ $=$ 0.05 K \cite{Onimaru_2010}. The coexistence of the AFQ order and the superconducting state suggests that the quadrupole fluctuations work as glue for superconducting electron pairs.
The AFQ order readily collapses by 5\% La substitution for the Pr ion, where the two-fold degenerated doublet is split by symmetry lowering of the cubic Pr site due to the atomic disorder, leading to a singlet ground state with no degrees of freedom as described in Supplementary Material \cite{Matsumoto_2015}.
Furthermore, the magnetic specific heat divided by temperature $C_{\rm m}{/}T$ exhibits  $-$ln$T$ dependence and $\rho(T)$ shows an upward curvature in a moderately wide temperature range
 above $T_{\rm Q}$ \cite{Onimaru_2016}.
These temperature dependences are consistent with those calculated by the two-channel Anderson lattice model, suggesting the formation of a quadrupole Kondo lattice \cite{Tsuruta_1999,Tsuruta_2015}. 
Moreover, a manifestation of a magnetic-field induced FL state hinted at a NFL-FL crossover due to formation of an electronic composite order in the quadrupole Kondo lattice \cite{Hoshino11,Hoshino13,Hastatic,Dyke_2018}.
If this is the case, a single-site quadrupole Kondo effect could manifest itself when the Pr$^{3+}$ ions in PrIr$_2$Zn$_{20}$ are diluted with rare-earth elements without $4f$ electrons.

In this Letter, we focus on a diluted Pr$^{3+}$ system Y$_{1-x}$Pr$_x$Ir$_2$Zn$_{20}$ showing single-site NFL behaviors in both $C_{\rm m}$ and $\rho$ at temperature below 3 K.
In fact, the $C_{\rm m}/T$ data for $x < 0.05$ exhibit $-$ln$T$ dependence down to the lowest temperature 0.08 K. 
Moreover, $\rho (T)$ shows upward curvature at $T$ $<$ 2 K.
Both sets of data $C_{\rm m}/T$ and $\rho$ per Pr ion are well scaled as a function of $T/T_{\rm 0}$ with the characteristic temperature $T_{0}$.
The temperature dependences agree with the forms calculated by the single-site quadrupole Kondo model with on-site strong coupling between the localized quadrupole and the two-channel conduction bands \cite{Cox_1998}.

Single-crystalline samples of Y$_{1-x}$Pr$_x$Ir$_2$Zn$_{20}$ were grown by the Zn-self flux method using high-purity elements: Y ($99.99\,\%$), Pr ($99.99\,\%$), Ir ($99.9\,\%$), and Zn ($99.9999\,\%$).
The atomic compositions were determined by the wave-length dispersive electron-probe microanalysis (EPMA) using a JEOL JXA-8200 analyzer.
However, the resolution for the Pr composition was not high enough for the range $x < 0.05$.
The value of $x$ was therefore estimated by comparing the magnetization values measured at $T = 1.8$ K and $B =1$ T with those calculated using the CEF level scheme for PrIr$_2$Zn$_{20}$ \cite{Iwasa_2013}. 
The electrical resistance was measured by a standard AC four-probe method in a laboratory built system with a Gifford-McMahon-type refrigerator between 3 and 300 K and a commercial Cambridge mFridge mF-ADR/100s adiabatic demagnetization refrigerator from 0.08 to 3 K.
The electric current was applied along the [100] direction.
The specific heat was measured by the thermal relaxation method using a Quantum Design physical property measurement system (PPMS) between 0.4 and 300 K and a laboratory built system with the adiabatic demagnetization refrigerator from 0.08 to 0.5 K.

\begin{figure}
\center{\includegraphics[width=8cm]{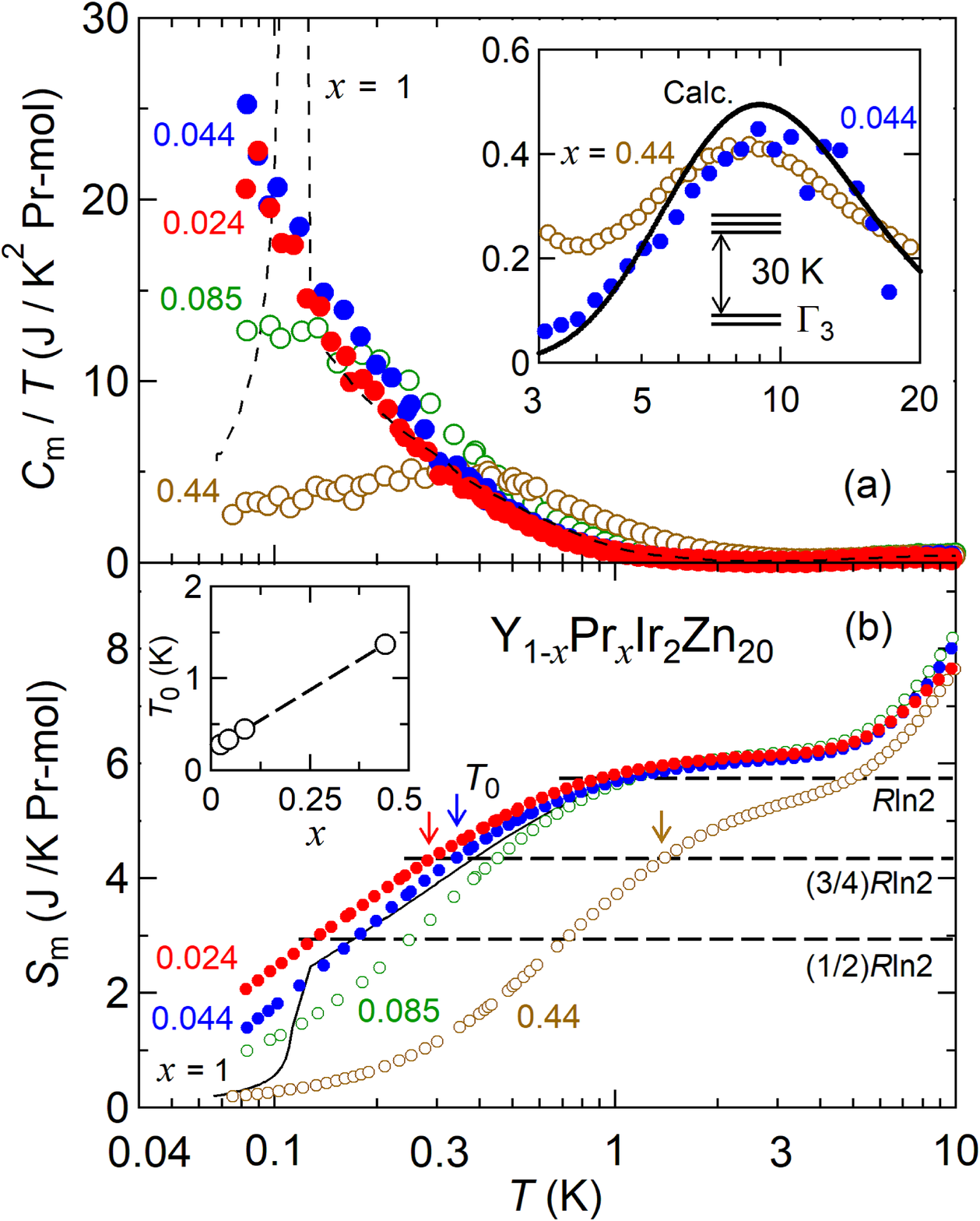}}
\caption{Fig.1 (Color online) (a) Temperature variations of the magnetic specific heat divided by temperature $C_{\mathrm{m}}/T$ per Pr mol for Y$_{1-x}$Pr$_x$Ir$_2$Zn$_{20}$.
The (black) dashed curve shows the data for $x$ $=$ 1 \cite{Onimaru_2011}.
The inset shows $C_{\mathrm{m}}/T$ for $x = 0.044$ and 0.44.
The (black) solid curve is calculated with a doublet-triplet two-level model with the energy gap of 30 K.
(b) Temperature dependences of magnetic entropy $S_{\mathrm{m}}$.
The (black) thin curve is that for $x = 1$ \cite{Onimaru_2016}.
Characteristic temperatures $T_0$ shown with the arrows are determined as the temperatures where $S_{\mathrm{m}}$ reaches (3/4)$R$ln2.
The inset shows $T_0$ vs $x$. }
\end{figure}

The data of $C_{\mathrm{m}}/T$ per Pr mol for Y$_{1-x}$Pr$_x$Ir$_2$Zn$_{20}$ with $x = 0.044$ and 0.44 are plotted in the inset of Fig. 1(a), where the $C_{\rm m}$ data are adopted from ref. \cite{Yamane_2017}.
The plots of $C_{\mathrm{m}}/T$ exhibit a maximum at around 9 K, which moderately agrees with the calculation for the CEF scheme of  doublet and triplet separated by 30 K as shown with the solid curve.
Since the magnitudes of $C_{\rm m}$ and the maximum temperature are same as in PrIr$_2$Zn$_{20}$ $(x =1)$ \cite{Onimaru_2011},
the Pr ions in the diluted system retain the $\Gamma_3$ doublet ground state with the same energy separation.

The main panel of Fig. 1(a) shows $C_{\mathrm{m}}/T$ for samples with $x \leq 0.44$ and $x = 1$ \cite{Onimaru_2016}.
On cooling below 0.4 K, all the data gradually increase.
At lower temperatures, the saturating behavior for $x = 0.085$ is reproduced by the calculation with a random two-level (RTL) model \cite{Yamane_2017}, in which the $\Gamma_3$ doublet is split randomly by the atomic disorder in the Pr sublattice.
On the other hand, $C_{\mathrm{m}}/T$ of more diluted samples with $x = 0.024$ and 0.044 does not saturate even at $T$ $<$ 0.3 K.
Instead, the $-$ln$T$ behavior continues to the lowest temperature of 0.08 K.

The magnetic entropy $S_{\mathrm{m}}$ per Pr ion is plotted in Fig. 1(b). 
For $x$ $=$ 0.085 and 0.44, the data of $C_{\mathrm{m}}/T$ approach almost constant values on cooling below 0.1 K as shown in Fig. 1(a). Thereby, we estimated $S_{\mathrm{m}}$ by integrating the data of $C_{\mathrm{m}}/T$ on the assumption that it remains the value at the lowest temperature of 0.08 K down to $T$ $=$ 0.
On the other hand, for $x = 0.024$ and 0.044, since $C_{\mathrm{m}}/T$ follows the $-$ln$T$ dependence to the lowest temperature of 0.08 K, we determined the offset of $S_{\mathrm{m}}$ for $T$ $\le$ 0.08 K so that the magnitude above 4 K agrees with the value for $B$ =4 T, where the full magnetic entropy of the ground state doublet assumed to be released by split of it due to the Zeeman effect through the excited magnetic triplet.
The data of $S_{\mathrm{m}}$ in the magnetic fields of $B$ $=$ 4 and 6 T and the calculation with the CEF parameters are presented in Fig. S1 of Supplemental Material.

With increasing $x$ from 0.024 to 0.44, the curve of $S_{\mathrm{m}}$ shifts to higher temperatures.
Now, we define a characteristic temperature $T_0$ as the temperature where $S_{\mathrm{m}}(T)$ reaches (3/4)$R$ln2.
This definition of $T_0$ is the same as that of the Kondo temperature $T_{\mathrm{K}}$ for the impurity quadrupole Kondo model \cite{Sacramento_1991,Cox_1998}.
This method gives $T_0$ for $x =0.024$ as 0.28 K, which value monotonically increases to 1.37 K for $x = 0.44$ as seen in the inset of Fig. 1(b).
Note that $C_{\mathrm{m}}/T$ data for $x < 0.05$ exhibit $-$ln$T$ behavior at $T \leq T_0$, suggesting that $-$ln$T$ behavior could result from the active quadrupoles of  the non-Kramers doublet.

\begin{figure}
\begin{center}
\center{\includegraphics[width=9cm]{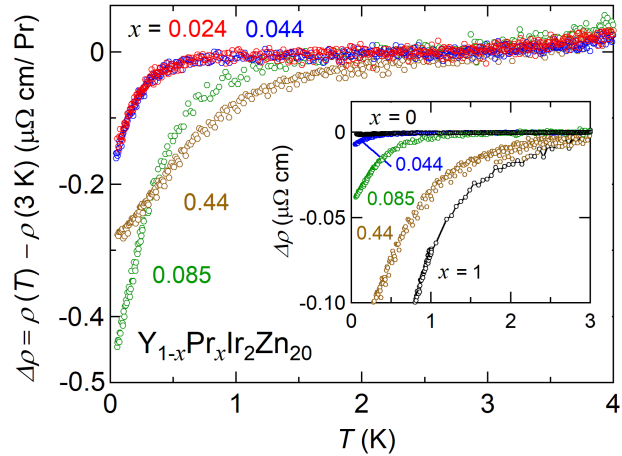}}
\caption{Fig.2 (Color Online) Temperature dependence of the differential electrical resistivity $\mathit{\Delta} \rho(T) = \rho(T) - \rho(3\: \mathrm{K})$ normalized by the Pr content $x$ for $x = 0.024, 0.044, 0.085, 0.44$ and 1. The magnetic field of 0.5 T was applied to kill the superconductivity. The inset shows the raw data of $\mathit{\Delta} \rho(T)$.}
\label{rho}
\end{center}
\end{figure}

\begin{figure}
\begin{center}
\center{\includegraphics[width=9cm]{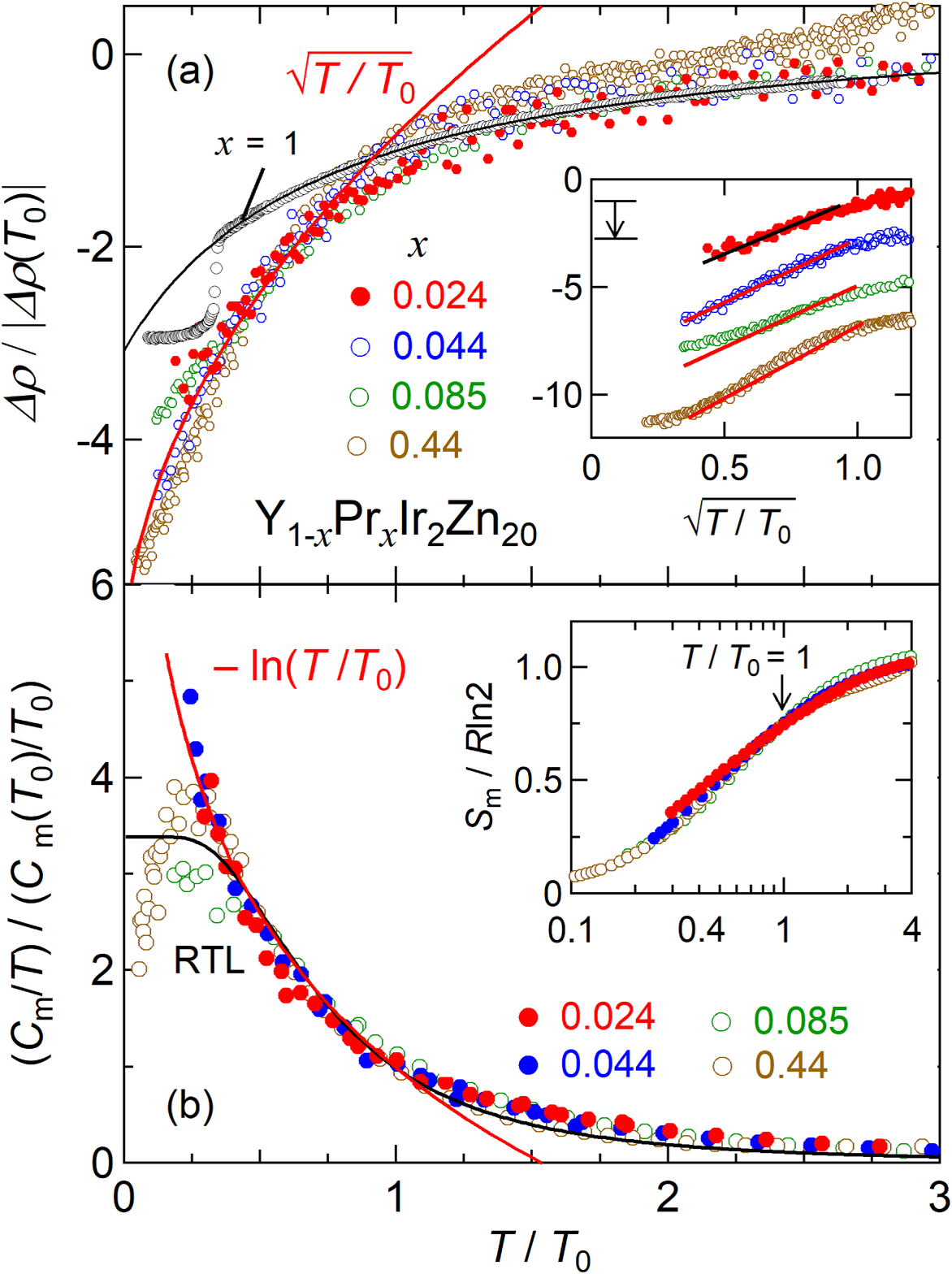}}
\caption{Fig.3 (Color online) Scaling plots of (a) $\mathit{\Delta} \rho$ and (b) $C_{\mathrm{m}}/T$ which are divided by the values at the characteristic temperature $T_0$ determined from the temperature variations of $S_{\mathrm{m}}$.
(Red) solid curves of $\sqrt{T}$ (a) and $-$ln$T$ (b) are predicted by the single-site quadrupole Kondo model \cite{Cox_1998}.  (Black) curves of  (a) and (b) are calculated with the quadrupole Kondo lattice model \cite{Tsuruta_1999, Tsuruta_2015} and the random two-level model \cite{Anderson_1972, Phillips_1972}, respectively. Inset: (a) $\mathit{\Delta} \rho/|\mathit{\Delta} \rho(T_0)|$ data as a function of $\sqrt{T/T_0}$, which are vertically offset for clarity. (b) $S_{\mathrm{m}}$ data as a function of $T/T_0$.}
\label{scale}
\end{center}
\end{figure}

The inset of Fig. \ref{rho} shows the temperature variations of the differential electrical resistivity from the value at 3 K, $\mathit{\Delta} \rho(T) = {\rho}(T) - {\rho}(3 $ K$)$.
Since the phonon contribution is fully suppressed at $T$ $<$ 3 K as referred to ${\rho}$$(T)$ of the nonmagnetic counterpart YIr$_2$Zn$_{20}$ ($x$ $=$ 0) \cite{Yamane_2017}, the electronic contribution can be extracted from the plot of $\mathit{\Delta} \rho(T)$.
To obtain $\rho(T)$ in the normal state for $x =0$, the superconducting transition at 0.12 K (not shown) was killed by the application of magnetic field of 0.5 T along the [100] direction.
Then, $\mathit{\Delta} \rho$ for $x = 0$ stays at a constant below 3 K.
The downward deviation in $\mathit{\Delta} \rho$ with an upward curvature becomes significant with increasing $x$ above 0.044, indicating that the upward dependence arises from the scattering by the Pr ions.
The main panel of Fig. \ref{rho} shows $\mathit{\Delta} \rho(T)$ normalized by $x$.
The normalized data for all samples exhibit upward curvature on cooling down to 0.08 K.
Especially, the agreement of the $\mathit{\Delta} \rho$ data normalized by $x$ for the two samples with $x = 0.024$ and 0.044 indicate that the upward curvature is a single-site effect of the Pr$^{3+}$ ions.
The curves shift to higher temperatures with increasing $x$, which is consistent with the increase in $T_{0}$ estimated from the analysis of \ $S_{\rm m}$ as mentioned above.

Shown in Figs. \ref{scale}(a) and  3(b), respectively, are $\mathit{\Delta} \rho$ and $C_{\mathrm{m}}/T$ divided by the absolute values at $T_0$ as a function of $T/T_0$.
The $\mathit{\Delta} \rho/|\mathit{\Delta} \rho(T_0)|$ data for $x = 1$ follow the form (black curve) predicted by the quadrupole Kondo lattice model \cite{Onimaru_2016}.
On the other hand, all $\mathit{\Delta} \rho/|\mathit{\Delta} \rho(T_0)|$ data for 0.024 $\le$ $x$ $\le$ 0.44 agree well each other and follow $\sqrt{T}$ at $T/T_0 \leq 1$, which is consistent with the $T$-dependence expected from the single-site quadrupole Kondo theory \cite{Cox_1998}.
Note that the sign of the coefficient $A$ of $\sqrt{T}$ is positive, being opposite to that predicted for the weak coupling between the $f^2$ electrons and the conduction electrons.
We recall that the sign could be positive in the strong coupling limit \cite{Affleck_1995}.

As seen in Fig. \ref{scale} (b), the normalized data $(C_{\mathrm{m}}/T) / (C_{\mathrm{m}}(T_0)/T_0)$ for all $x$ agree in the range 0.5 $\le$ $T/T_0$ $\le$ 3.
In this temperature range, we found the scaling of $\mathit{\Delta} \rho$ as shown in Fig. \ref{scale} (a).
The good scaling of the two quantities $\rho$ and $C_{\mathrm{m}}$ indicates the NFL behaviors to originate from the same mechanism.
In the range $T/T_{0}$ $<$ 0.5, however, the data of  $C_{\mathrm{m}}/T$ strongly depend on $x$.
For $x < 0.05$, the $-$ln$T$ dependence holds down to the lowest temperature, in agreement with the prediction from the single-site quadrupole Kondo model \cite{Cox_1998}, as mentioned above.

At $x = 0.085$, the data of  $C_{\mathrm{m}}/T$ deviate from the $-$ln$T$ form and are saturated on cooling below $T/T_{0}$ $=$ 0.5, which can be explained by the RTL model \cite{Anderson_1972, Phillips_1972}, as shown with the (black) curve \cite{Yamane_2017}.
At the highest concentration $x =0.44$, a maximum at $T/T_0 = 0.3$ is followed by a significant decrease on cooling, whose behavior is likely to be a result of short-range correlation between quadrupoles.
The $\mathit{\Delta} \rho(T)$ data also become dependent on $x$ at $\sqrt{T/T_{0}}$ $<$ 0.5 as shown in the inset of Fig. \ref{scale} (a).
The $\sqrt{T}$ dependence of $\mathit{\Delta} \rho/|\mathit{\Delta} \rho(T_0)|$ for $x < 0.05$ holds down to the lowest temperature.
However, for $x =$ 0.085 and 0.44, the data deviate from the $\sqrt{T}$ dependence and reach constant values at $\sqrt{T/T_{0}}$ $<$ 0.5 ($T/T_{0}$ $<$ 0.25).
The concomitant deviation of both $C_{\mathrm{m}}/T$ and  $\mathit{\Delta} \rho$ from the curves calculated with single-site quadrupole Kondo model suggests that a singlet ground state could be formed by breaking the degeneracy of the equivalent conduction bands due to the symmetry lowering of the cubic Pr site or development of the inter-site interaction.

Here, it should be noted that the NFL behaviors of $\rho$ and $C_{\rm m}$ were observed for $T/T_{0}$ $<$ 1, although they are expected to manifest themselves at much lower temperature range compared to the Kondo temperature $T_{\rm K}$ in the two-channel Kondo model \cite{Sacramento_1991,Cox_1998}. 
The disagreement suggests that the characteristic temperature $T_{0}$ determined in the present work is not equal to $T_{\rm K}$. 
In fact, the measured magnetic entropy changes more rapidly with $T/T_{0}$ than the calculation with respect to $T/T_{\rm K}$. 
Thereby, a characteristic temperature corresponding to $T_{\rm K}$ in the present system must be much lower than $T_{0}$.
This discrepancy between $T_{0}$ and $T_{\rm K}$ probably results from collapse of the equivalency of the two-channel conduction bands by symmetry lowering due to a small amount of atomic disorder.
Calculation using the two-channel Kondo model predicted that the magnetic entropy shifts to higher temperatures in magnetic fields since the two-channel conduction bands becomes inequivalent \cite{Sacramento_1991}.

Now, let us discuss the residual entropy of $0.5R$ln2 at $T$ $=$ 0 that is the hallmark of the single-site quadrupole Kondo state \cite{Cox_1998} and fictional Majorana fermions \cite{Emery_1992, Alicea_2012}.
To confirm the existence of the residual entropy, we measured the specific heat in magnetic fields, since the residual entropy has to be fully released by breaking the equivalency of the two-channel conduction bands in the magnetic fields as explained above. 
In fact, as shown in Fig. S1 of Supplementary Material, the obtained $S_{\rm m}$ data for $x$ $=$ 0.044 in $B$ $=$ 4 T falls to zero on cooling at around 0.1 K, indicating that the magnetic entropy of 0.25$R$ln2 exists at least for $T$ $<$ 0.08 K as shown in Fig. 1(b) and the inset of Fig. 3(b).
On the other hand, even for $x=0.024$, $S_{\rm m}$ at 0.08 K is about $0.3R$ln2 that is less than $0.5R$ln2.
The shortage of entropy suggests that the $\Gamma_3$ doublet ground state is randomly split by the atomic disorder in the Pr-Y sublattice, where the equivalency of the two-channel conduction bands is broken through the $c$-$f$ hybridization. 
The effect of the inequivalent bands on the rapid release of $S_{\rm m}$ on cooling was also suggested above.
Another possible origin for the shortage of the entropy would be hyperfine interaction of the quadrupoles of the 4$f^2$ electrons with those of $^{141}$Pr nucleus.
In fact, enhanced nuclear magnetism of the $^{141}$Pr nucleus were observed in Pr-based metallic systems \cite{Kubota_1983}. In PrRu$_4$P$_{12}$, the Schottky-type specific heat arising from formation of $4f$-electron-nuclear hyperfine-coupled multiplets was reported \cite{Aoki_2011}.

In summery, we have reported the definite observation of single-site NFL behaviors in both $\rho(T)$ and $C_{\rm m}(T)$ for an isolated quadrupole-active state of Pr$^{3+}$ in Y$_{1-x}$Pr$_x$Ir$_2$Zn$_{20}$.
The data of $C_{\rm m}/T$ and $\mathit{\Delta}{\rho}$ normalized by, respectively, the values at the characteristic temperature $T_{0}$, are well scaled as a function of  $T/T_0$.
Moreover, on cooling below $T_0$, the $C_{\rm m}/T$ data for $x < 0.05$ logarithmically diverge and $\rho (T)$ follows $1+A\sqrt{T}$ with the positive coefficient of $A$.
These findings imply that the single-site NFL behaviors are originated from the single-site quadrupole Kondo effect.
If the on-site coupling between the localized quadrupole and two-channel conduction bands is efficiently strong, the elastic modulus would exhibit $-$ln$T$ dependence \cite{Cox_1998}.
Moreover, insights into single-site quadrupole fluctuations would be given by microscopic measurements such as nuclear magnetic/quadrupole resonance and/or inelastic neutron scattering.

The authors would like to thank H. Kusunose, K. Izawa, S. Hoshino, J. Otsuki, A. Tsuruta, Y. Tokiwa, R. J. Yamada, and G. B. Park for helpful discussion. The authors also thank Y. Shibata for the electron-probe microanalysis carried out at N-BARD, Hiroshima University. The measurement of the specific heat with the PPMS and the mFridge mF-ADR 100s refrigerator were performed at N-BARD, Hiroshima University. This work was financially supported by Grants-in-Aid from MEXT/JSPS of Japan, Nos. JP26707017, JP15KK0169, JP15H05886, and JP16H01076 (J-Physics).

\section{Supplementary Materials}
\subsection{\label{sec:level1}Magnetic entropy in magnetic fields}

The specific heat measurements of Y$_{1-x}$Pr$_x$Ir$_2$Zn$_{20}$ for $x =0.044$ have been performed at temperatures down to 0.1 K and in the magnetic fields of $B$ = 4 and 6 T along the [100] direction.
It {was} expected that the possible residual entropy {was} released at a higher temperature because of splitting of the ground state doublet by the Zeeman effect through off-diagonal elements to the excited triplet.

The experimental and calculated curves of the magnetic entropy $S_{\mathrm{m}}$ are shown in Fig. S1(a) and (b), respectively.
The curves in Fig. S1(a) were estimated by integrating the data of the magnetic specific heat divided by temperature $C_{\mathrm{m}}/T$ from the lowest temperature of about 0.1 K.
To estimate the $4f$ contribution to the total specific heat $C$ in magnetic fields, the nuclear and phonon contributions have been subtracted by following the procedure reported in Ref.\cite{Onimaru_2016}.
Here, we used the values of the isothermal magnetization for PrIr$_2$Zn$_{20}$ measured at $T$ = 0.045 K as the total angular momentum $\bm{J}$ of 4$f$ electrons multiplied by the Lande's $g$ factor.
Since the phonon contribution was unchanged to the magnetic field, it was subtracted by the way described in Ref. \cite{Yamane_2017}.

The calculated curves shown in Fig. S1(b) were obtained by considering the crystalline electric field (CEF) parameters of $W = -$1.22 and $X$ = 0.53 \cite{Iwasa_2013} and the Zeeman effect. Since the ground state doublet is not split in $B$ $=$ 0 by this calculation, the calculated curve for $B$ $=$ 0 is vertically offset by $R$ln2.

In order to evaluate the offset for measured $S_{\rm m}$ at $B$ = 0, we compared the experimental data to the calculated curve for $T$ $>$ 2 K. Thereby, the measured $S_{\rm m}$ for $B$ = 0 is vertically shifted by {1.4} J$/$K Pr-mol as shown with the black closed circles in Fig. S1(a).
The obtained experimental curves shifts to higher temperatures with increasing magnetic fields. 
\begin{figure}
\center{\includegraphics[width=8cm]{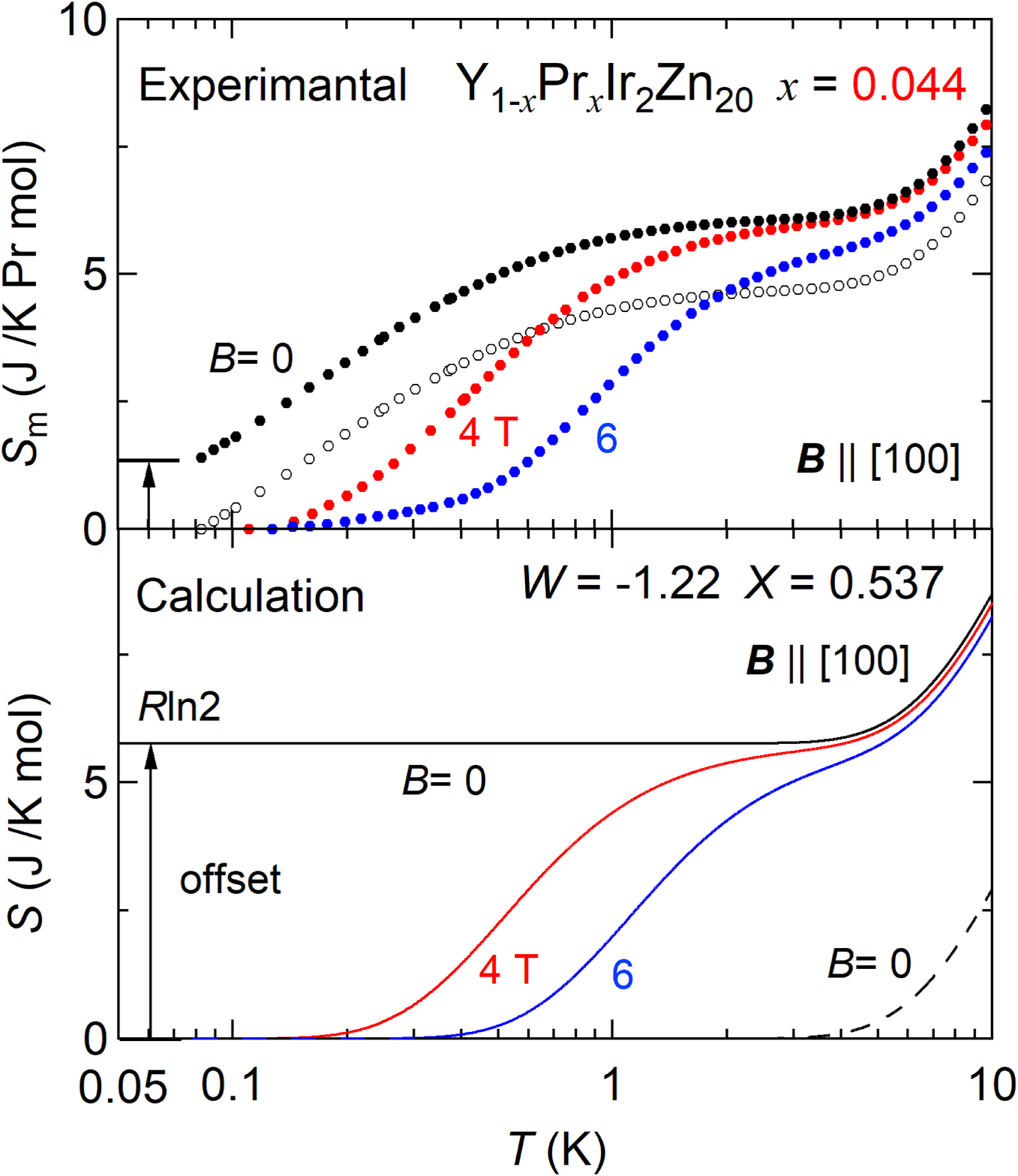}}
\caption{Fig. S1. (Color online) Temperature dependence of the magnetic entropy in the magnetic field of $B$ $=$ 0, 4, and 6 T applied along the [100] direction. (a) Experimental data. The data for $B$ = 0 are offset vertically by comparing the data with the calculation. (b) Calculations by using the CEF parameters of $W$ = -1.22 and {$X$} = 0.537 obtained by the inelastic neutron scattering experiments for PrIr$_2$Zn$_{20}$ \cite{Iwasa_2013}. Since the ground state doublet is not split in $B$ $=$ 0 by this calculation, the calculated curve is vertically offset by $R$ln2.}
\end{figure}

\begin{figure}[h]
\begin{center}
\center{\includegraphics[width=7cm]{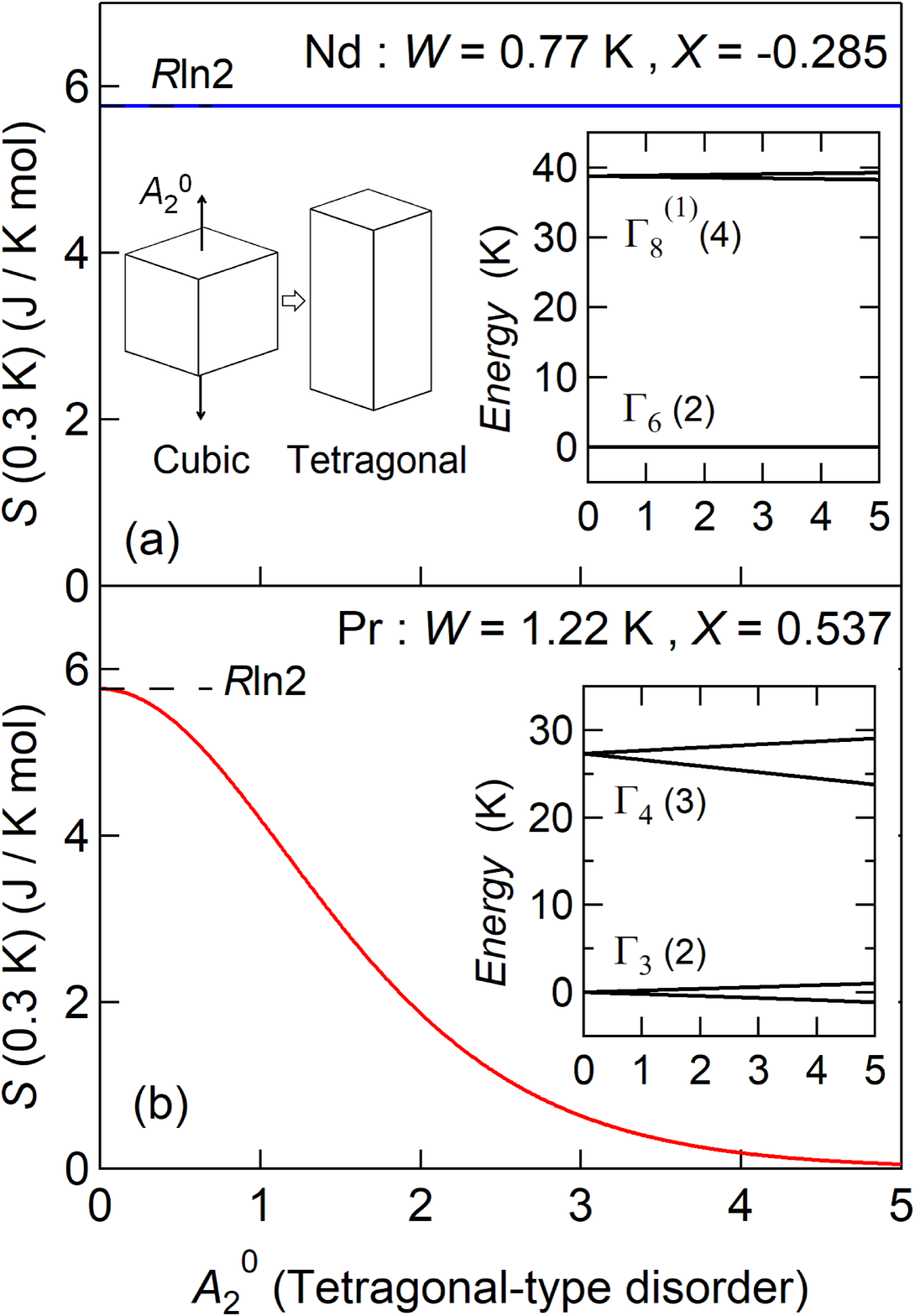}}
\caption{Fig. S2. Effect of symmetry lowering on magnetic entropy at $T$ = 0.3 K of (a) a Kramers doublet for a Nd$^{3+}$ ion with $4f^3$ configuration and (b) the non-Kramers doublet for the Pr$^{3+}$ ion. The insets show the eigenvalues of the CEF levels with the second-order tetragonal CEF parameter $A_2^0$. {The values in the parentheses following the symbols of the irreducible representations describe the degeneracy of the energy level at $A_2^0 = 0$.} }
\end{center}
\end{figure}

\subsection{\label{sec:level2}Crystalline electric field analyses considering tetragonal-type distortion}

Figure S2 shows effect of symmetry lowering on entropy of a Kramers doublet for a Nd$^{3+}$ ion with $4f^3$ configuration and the non-Kramers doublet for the Pr$^{3+}$ ion at $T$ = 0.3 K, which is much lower than the crystal-field energy gap of a few 10 K between the ground state doublets and the first excited states.
The Hamiltonian is presented below;
\begin{equation}
	\mathcal{H}_{\mathrm{CEF}} = A_2^0 \langle r^2 \rangle \alpha_J O_2^0 + W\Bigl[X \frac{O_4^0 + 5 O_4^4}{F(4)}+(1-\left|X\right|)\frac{O_6^0 - 21O_6^4}{F(6)}\Bigr],
\end{equation}
where $A_2^0$ is a second-order CEF parameter, which induces tetragonal-type disorder.
$\alpha_J$ is the Stevens factor and $\langle r^2 \rangle$ is the average of square of the orbital radius of the $4f$ electrons.
$W$ and {$X$} are CEF parameters and $O_n^m$ are the Stevens equality operators.
The values of $F(4)$ and $F(6)$ are 60 and 1260 for Pr$^{3+}$ and 60 and 2520 for Nd$^{3+}$, respectively.
Here, we set the CEF parameters as $W$ = 0.77 K and {$X$} = $-$0.285 for Nd$^{3+}$ \cite{Yamane_2017_2} and $W$ = 1.22 K and {$X$} = 0.537 for Pr$^{3+}$ \cite{Iwasa_2013}.

In the case of the Kramers doublet of Nd$^{3+}$ as shown in the {upper} panel, the entropy stays at $R$ln2 even if $A_2^0$ is increased from zero.
It is because the Kramers doublet is not split by any symmetry lowering as long as the time reversal symmetry is kept.
If the time reversal symmetry is broken, e.g. by a magnetic field or a magnetic exchange interaction, the entropy of the doublet should be released to form a spin glass state of localized magnetic moments and further to a magnetically ordered state.

On the other hand, as shown in the lower panel, the entropy of $R$ln2 in the non-Kramers doublet of the Pr ion {markedly} decreases with increasing $A_2^0$, indicating the entropy is released by the symmetry lowering.
{As a result}, the ground state {changes to} the nonmagnetic singlet with no quadrupolar degrees of freedom, that is to say, neither a quadrupole order nor a quadrupole glass state must be the ground state.
This situation is realized in the present system Y$_{1-x}$Pr$_x$Ir$_2$Zn$_{20}$ for $x$ = 0.085; the $C_{\mathrm{m}}/T$ data approach the constant values on cooling below 0.1 K and the temperature variations can be well reproduced by the random two-level model as shown in Eq. (2) and the (black) solid curve in Fig. 3(b).
\begin{equation}
	C_{\mathrm{RTL}}(T)= Nk_{\mathrm{B}} \int_0^\infty  n(E) (\frac{E}{k_{\mathrm{B}}T})^2 \frac{e^{-\frac{E}{k_{\mathrm{B}}T}}}{(1+e^{-\frac{E}{k_{\mathrm{B}}T}})^2} dE
\end{equation}
Here, the density of states of the ground state doublet $n(E)$ is assumed to be $1/\mathit{\Delta}$ in the energy range from zero to $\mathit{\Delta}$.
In the case for $x$ = 0.085, the ground state is the non-magnetic singlet with no quadrupolar degrees of freedom.

\end{document}